\documentclass[pre,twocolumn,a4paper,showpacs]{revtex4}
\usepackage{amssymb}
\usepackage{epsfig}
\begin{document}

\title{Preservation of information in a prebiotic package model}

\author{Daniel A. M. M. Silvestre and Jos\'e F. Fontanari}
\affiliation{Instituto de F\'{\i}sica de S\~ao Carlos,
  Universidade de S\~ao Paulo,
  Caixa Postal 369, 13560-970 S\~ao Carlos, S\~ao Paulo, Brazil}
                              

\begin{abstract}
The coexistence between different informational molecules has been the preferred
mode to circumvent the limitation posed by
imperfect replication on the amount of information stored by each of these molecules.  Here we reexamine a classic 
package model in which distinct information carriers or templates are forced to coexist
within vesicles, which in turn  can proliferate freely through binary division.
The combined dynamics of  vesicles and templates is described by
a multitype branching process which allows us to write 
equations for the average number of the different types of vesicles  as well as
for their extinction probabilities. The threshold phenomenon associated with the 
extinction of the vesicle population is studied quantitatively using finite-size scaling
techniques. We conclude that the resultant coexistence is too frail in the presence of
parasites and so  confinement of  templates in vesicles without an explicit mechanism of
cooperation does not resolve the information crisis of prebiotic evolution.

\end{abstract}

\pacs{87.10.+e, 87.23.Kg, 02.50.Ey, 02.70.Uu}

\maketitle

\section{Introduction}

The information crisis in prebiotic or chemical evolution stems essentially from two observations: (i) the
length of a replicating polymer  (i.e., RNA-like template) is limited by the replication accuracy 
per nucleotide \cite{Eigen:71}, and (ii) templates that differ significantly  from each other 
cannot coexist in a purely competitive setup \cite{Swetina:82}. 
Realistic estimates of the error rate of primitive replication mechanisms
predict a too scanty information content per template - less than 100 nucleotides -
to permit the complete codification of the mechanism in just one template.
Currently an operative replication mechanism requires at least three
basic sets of different reactions (initiation, elongation and termination) \cite{Elena:06}, 
thus the primitive information
integrator systems must have shared the necessary information in a number of distinct templates. Yet, attainment
of template coexistence in a plausible prebiotic scenario is still a highly controversial issue. 

An attractive solution to this crisis is the hypercycle, a cyclic reaction scheme in which each  replicating polymer
aids in the replication of the next one, in a regulatory cycle closing on itself \cite{Eigen:78}.
This scheme requires that the primordial replicators functioned 
both as templates and replicases (i.e., catalysts for replication), a prospect
confirmed   by the  discovery of the  
catalytic activity of RNA  in the early 1980s \cite{Altman:90,Cech:90}. However,
the key assumption that each replicator has two separate functions, namely a replicase for 
the next member of the hypercycle and a target for the previous 
member, encountered strong criticism \cite{Bresch:80,Maynard:79}. In fact, natural selection 
can make each element of the hypercycle a better target for replication, but it cannot favor
the cooperative part of the scheme, i.e., to make the replicator a better replicase 
for other replicators.  Hence this  function is bound to  degenerate rapidly
as natural selection does not protect it against deletions and mutations that
create the so-called parasites -- molecules that do not
reciprocate the catalytic support they receive.  

Another proposal to resolve the problem of the coexistence between templates,  which is in 
line with the classical works on the origin of life \cite{Oparin:54,Fox:65},  is to enwrap the
replicators in isolated compartments or vesicles. It is presumed that
the information to 
code for a common replicase is shared among $d$ distinct template types  so that template
replication is feasible only if all template types coexist within a vesicle \cite{Niesert:81}. 
In this scheme the replicators play the template role only whereas the replicase role is taken on by
a protein -- the replicase. In addition, it is assumed that a vesicle splits into two daughters after a certain 
number of template copies are produced.
Alternatively, we may suppose that an effective  coupling among different template types  
comes about through a 
common metabolism which is ultimately  responsible for the survival and reproduction of the vesicle, and 
that  the functioning of this metabolism requires the contribution of all template types \cite{Szathmary:87}. 

It should be noted, however, that the proponents of the hypercycle have always acknowledged the essential function
of compartments,  particularly in the  evaluation of the translation products of the
information coded in the templates \cite{Eigen:80} (see \cite{Matsuura:02} for the \textit{in vitro} 
realization of this idea). In reply, the advocates of the so-called package models 
observe that 
once the replicators are put into compartments, the hypercyclic organization is dispensable \cite{Bresch:80}.

So far practically all package models proposed to investigate template coexistence  (see, e.g., 
\cite{Szathmary:87,Czaran:00,Zintzaras:02,Silvestre:05,Fontanari:06})  assume  that the vesicles 
proliferate or divide with a rate that depends on their template compositions.
This assumption can be interpreted as a group selection pressure acting  at 
the vesicle level to favor  vesicles of a particular makeup.   Despite 
years of intensive research on vesicle dynamics, however,  there is no experimental
evidence that the vesicle fission rate  could depend on the nature of the chemicals confined inside it
\cite{Hanczyc:03,Chen:04}. Hence a more conservative stand is to admit that vesicle fission is triggered by  
the total concentration of the confined templates rather than by their individual proportions. 
(Of course, the total template concentration does depend on the existence of the replicase and hence
on the presence of all $d$ template types.) Interestingly, in their seminal work Niesert et al.\ 
take this cautious position, but to avoid the unbounded growth of the vesicle population, 
they discard supernumerary vesicles  according to an 
arbitrary prospective value which essentially gauges the odds of a vesicle to leave viable descendents 
\cite{Niesert:81}. 
Then the resulting model becomes 
very similar to the group selection  models mentioned above.

In this paper we propose and study analytically a variant of the original package model of
Niesert et al. where the number of vesicles is unbounded and no fitness or prospective values
are assigned to the vesicles. In particular, we derive a recursion equation for the average number of 
vesicles with a given template composition, and a set of equations for the extinction probability
of the distinct vesicle types. Our results indicate that an important conclusion of the original
work -- high values of the replicase processivity compromises the viability of the population
in the presence of parasites -- is probably an artifact of sampling the vesicles 
according to a prospective value. In this line, we use finite size scaling to
show how a strategy of discarding supernumerary vesicles at random can efficiently
recover the analytical results.

%
\section{Model}\label{sec:model}
%

We follow  the original package  model proposed by Niesert et al.\ \cite{Niesert:81}  
and consider a metapopulation composed of a variable number of vesicles, 
each of  which encloses a certain number of templates. There are $d$ distinct functional 
types of templates $l=1,\ldots,d$ and a non-functional type $l=0$,  termed parasite, which has 
an impaired function but an unchanged replication rate. Due to imperfect replication, functional 
templates mutate to parasites with probability $u$. Back mutations as well as mutations between 
functional templates are neglected. To be consistent with the conjecture that all templates 
display identical targets to the replicase since they  derive from a common ancestor (most likely
were members of a same quasispecies \cite{Eigen:71}), the 
replication rate is assumed to be the same for all templates (including the parasite). 
Here we do not contemplate two additional processes allowed for in the original model, namely, 
the possibility of mutation to lethal genes or the possibility of accidents. 
Both actions prompt  the immediate  demise of the vesicle. 

The life cycle (i.e., one generation) of the metapopulation  
comprises  three events - template replication, vesicle fission and vesicle extinction - 
that take place in this order. In this contribution we modify the first two events 
in order to produce an analytical formulation of the metapopulation dynamics. In particular,
we assume that the quantity of templates confined in each vesicle before replication is fixed to 
a certain value $\Lambda$. Template replication doubles this number but then fission of the mother vesicle
into two daughters of identical size restores it to the original value. Hence $\Lambda$ can
be interpreted as the number of  replicated molecules between two vesicle fissions, which is essentially 
the processivity of the replicase, i.e., the number of template copies the replicase 
can produce in a unit of time, taken here as the time between two consecutive fissions. 

In contrast,
in the formulation of Niesert et al.\ \cite{Niesert:81} the two daughter vesicles can have different sizes $s =0,\ldots,S$ and
$S- s$, with $s$ distributed
by the   binomial distribution 
$\left ( \begin{array}{c} S \\ s \end{array} \right ) 2^{-S}$ where $S$ is the size of the mother vesicle after
template replication, so that the
number of templates can vary among the vesicles. The processivity $\Lambda$ of the replicase, however, is
the same for all vesicles and so, in the average, the number of templates within each vesicle equals $\Lambda$.
Thus, essentially, our formulation  neglects fluctuations in the number of templates
inside the vesicles. As we will show in Sec.\ \ref{sec:Disc}, these variations in the modeling of
the vesicle dynamics do not change qualitatively the main results of the model.

In our model, the composition of each vesicle is fully characterized by the  
vector $\vec{k} = \left ( k_0, \ldots,k_d  \right )$, where the entries
$k_l$  yield the number of templates of type  $l=0,\ldots,d$
in the vesicle and satisfy the constraint $\sum_{l=0}^d k_l = \Lambda$. So there are
exactly $ \mathcal{N}_T = \left ( \begin{array}{c} \Lambda + d \\ \Lambda \end{array} \right )$ distinct
types of vesicles - the number of compositions of $\Lambda$ into $d+1$ parts. To
keep track of all vesicle types we use a combinatorial algorithm to generate and label 
those compositions \cite{Ninjenhuis:78}.

The  more restrictive assumption of the model is probably the choice of equal
replication rates for the distinct types of functional templates as well as for the parasites. 
The supposition that the parasite and the functional classes have equal replication rates is 
plausible since
a parasite is essentially a functional template whose activity was impaired by a mutation in the region 
coding for a piece  of the replicase.  In any event, allowing the parasites to 
replicate faster or slower than the functional templates has an effect similar to that of increasing
or decreasing the mutation probability $u$.
The choice of different replication rates for the
functional templates, however, has drastic consequences in the limit of large $\Lambda$, where the intra-vesicle dynamics 
becomes deterministic: the more efficient template type drives the other functional templates to
extinction, thus preventing coexistence even in the absence of parasites. 
In fact, in a class of models where the number of vesicles is fixed and very large, there is a limiting 
value of $\Lambda$ above which template coexistence is impossible \cite{Silvestre:05,Fontanari:06}.
We note, however, that a more realistic
scenario would allow the  replication rates of the functional templates to vary under the pressure of
natural selection. Since only the exact balancing of those rates guarantees coexistence (and so
survival) for large $\Lambda$,
one expects the selection of this ideal symmetric setting. This reasoning  supports
the assumption of equal replication rates for the functional templates.

\subsection{Template replication}

We assume that the replication of the $\Lambda$ templates encaged in a
vesicle follows a Wright-Fisher process in 
which the $\Lambda$ offspring are chosen in parallel \cite{Ewens:04}. Admitting equal replication rates  and unidirectional mutation
to the parasite class, 
the probability that a set of templates $k_0, \ldots, k_d$ 
 produces the set of offspring $i_0,\ldots, i_d$ 
is given by the multinomial distribution
\begin{equation}\label{multi_R}
R \left ( \vec{i} \mid \vec{k} \right ) = \frac{\Lambda!}{i_0! \ldots i_d!} \left [ w_0 + u \left ( 1 -w_0 \right ) \right ]^{i_0}
\prod_{l=1}^d \left [ w_l \left ( 1 - u \right ) \right ]^{i_l} .
\end{equation}
where $w_l = k_l/\Lambda$  for $l=0, \ldots,d$ so that $\sum_{l=0}^d w_l = 1$. The interpretation
of Eq.\ (\ref{multi_R})  is straightforward. On the one hand, the probability of producing a functional offspring of type
$l$ is given by  the probability of choosing a template of the same type, $w_l$, times
the probability that the copy produced is faithful, $1-u$. Parasites, on the other hand, are produced by unfaithful 
copies of functional templates with probability $u \left ( w_1 + \ldots + w_d \right )$ or
by copies of parasites themselves, with probability $w_0$. Once the template replication process is completed, we are left with a 
vesicle of size  $ 2 \Lambda $ and composition  $k_0 + i_0, \ldots, k_d + i_d$. This procedure is repeated for all vesicles in 
the metapopulation.

In the original model \cite{Niesert:81} the replication procedure is sequential rather
than parallel. For a given vesicle we choose a 
template at random (from those inside the vesicle) and make a copy of it. If 
the template is of a functional type then the copy will become a parasite with probability $u$. 
If the chosen template is a parasite then the copy will also be a parasite. Both template and 
copy (corrupted or not)  are returned to the vesicle and the process is repeated $\Lambda$  times, so
exactly $\Lambda$ new templates are added to the vesicle. This difference in the
modeling of the template replication process does not affect the results in any significant way.

\subsection{Vesicle fission}

The doubling of the size of the vesicles caused by the template  replication process  leads to the 
splitting of the vesicle in two daughters. Our simplifying assumption here is that the vesicle of size  $2 \Lambda$
splits into two vesicles of size $\Lambda$. 
The assignment of the  $\Lambda$ templates to one of the daughter vesicles is modeled by a process of sampling without 
replacement which is described by a multivariate hypergeometric distribution. Explicitly, given 
the composition of the mother vesicle after template replication $\vec{k} + \vec{i}$,  the probability that one of the 
daughter vesicles has composition  $m_0, \ldots, m_d$ is simply
\begin{equation}\label{hyper}
F \left ( \vec{m} \mid \vec{k} + \vec{i} \right ) = \frac{\prod_{l=0}^d \left ( \begin{array}{c} k_l + i_l \\ m_l \end{array} \right )}
{\left ( \begin{array}{c} 2 \Lambda  \\ \Lambda \end{array} \right )}
\end{equation}
with $m_l \leq k_l + i_l$ and $\sum_{l=0}^d m_l = \Lambda$.  
Of course, if one of the daughter vesicles is described by  $\vec{m}$, 
then the other will be  described by $\vec{k} + \vec{i} - \vec{m}$.
The random assortment of templates to the daughter vesicles  is the 
only mechanism responsible for the loss of the essential genes for survivorship, a phenomenon termed assortment load. The loss of
a functional template occurs when it is assigned to an inviable vesicle, i.e., a vesicle that does not contain
the complete set of functional templates.

As mentioned before, in the original model \cite{Niesert:81} the sizes of the daughter vesicles are binomially distributed random variables
and so some vesicles can become very large  since  what prompts vesicle fission is not its absolute size, but the
production of $\Lambda$ template offspring. This asymmetry in the fission process renders the population  more susceptible
to the presence of parasites (see Sec. \ref{sec:Disc}), but as already said, does not change qualitatively the results  of
the model. 

\subsection{Vesicle extinction}

The viability of a daughter  vesicle is guaranteed provided it has at least one copy of each functional template. 
Any vesicle lacking one of those templates is dismissed. Strictly, we do not need to assume that the inviable vesicles 
disappear from the metapopulation, but since the templates caged in those vesicles
are unable to replicate - their replicase is not codified for - there is no point to follow their evolution any further. 
We note that the total number of viable
vesicles  $\mathcal{N}_V = \sum_{k_0 \geq 0, k_l \geq  1}  \delta \left ( \sum_l^d k_l,\Lambda \right )$, where
$\delta (m,n)$ is the Kronecker delta, is 
simply $ \left ( \begin{array}{c} \Lambda \\ d \end{array} \right )$.   
To the leading order in $\Lambda$ both quantities $\mathcal{N}_V$ and $\mathcal{N}_T$
increase as $\Lambda^d$ and for large $\Lambda$ we 
find  $\mathcal{N}_V/\mathcal{N}_T = 1 - d^2/\Lambda + \mathcal{O} \left ( \Lambda^{-2} \right )$.

\subsection{Metapopulation dynamics} \label{sec:recursion}

>From the processes described above, it is clear that the size of  the metapopulation  
(i.e., the number of viable vesicles) can, in some cases, increase without bounds. 
Such unbounded growth renders  a direct simulation approach of the vesicle
population dynamics unfeasible, except for the few initial generations. To circumvent
this difficulty, here we derive a set of recursion equations for
the average number of vesicles of  type $\vec{m}$ at  generation $t$, denoted by
$\Phi_{t} \left ( \vec{m} \right )$. 

The basic idea is to  derive a transition matrix that connects the mother vesicle $\vec{k}$
with the two daughter vesicles $\vec{m}_a$ and $\vec{m}_b$. From Eq.\ (\ref{hyper}) we can
immediately write down the transition probability from the mother vesicle to the
first daughter,
\begin{equation}\label{G_a}
G_a \left ( \vec{m}_a \mid \vec{k} \right ) = \sum_{\vec{i}} F \left ( \vec{m}_a \mid \vec{k} + \vec{i} \right ) 
R \left ( \vec{i} \mid \vec{k} \right ) .
\end{equation}
The derivation of the transition probability from $\vec{k}$ to the second daughter $\vec{m}_b = \vec{k} + \vec{i} - \vec{m}_a$
is more involved because of the dependence on the intermediate states $\vec{i}$ which we ultimately
want to sum over, as done in Eq.\ (\ref{G_a}).  Given that the first daughter has composition $\vec{m}_a$, 
the probability that the second daughter has composition  $\vec{m}_b$ is 
\begin{equation}\label{H}
H \left ( \vec{m}_b \mid  \vec{m}_a \right )  = \frac{F \left ( \vec{m}_a \mid  \vec{m}_b + \vec{m}_a \right )}
{G_a \left ( \vec{m}_a \mid \vec{k} \right )}
R \left ( \vec{m}_b + \vec{m}_a - \vec{k} \mid  \vec{k} \right )  
\end{equation}
which is obtained by considering only the term  $\vec{i} = \vec{m}_b  + \vec{m}_a - \vec{k}$
in Eq.\ (\ref{G_a}) and properly normalizing. Clearly, the joint probability that
the daughter vesicles are of types $\vec{m}_a$ and $\vec{m}_b$ given that the mother vesicle is of type $\vec{k}$ is
simply
\begin{equation}\label{joint}
P_{ab} \left ( \vec{m}_a, \vec{m}_b \mid \vec{k} \right )  =  H \left ( \vec{m}_b \mid  \vec{m}_a \right )
G_a \left ( \vec{m}_a \mid \vec{k} \right ) ,
\end{equation}
and so the desired transition probability is 
\begin{eqnarray}\label{G_b}
G_b \left ( \vec{m}_b \mid \vec{k} \right ) & = & \sum_{\vec{m}_a}  
P_{ab} \left ( \vec{m}_a, \vec{m}_b \mid \vec{k} \right )\nonumber \\
& = & \sum_{\vec{i}} F \left ( \vec{i} \mid \vec{m}_b + \vec{i} \right ) 
R \left ( \vec{m}_b + \vec{i} - \vec{k} \mid \vec{k} \right ) , \nonumber \\
& &
\end{eqnarray}
where we have replaced the dummy index $\vec{m}_a$ by $\vec{i}$ to facilitate the comparison with Eq.\ (\ref{G_a}).
The transition matrices given by Eqs.\ (\ref{G_a}) and (\ref{G_b}) allow us to write a recursion equation for the
average number of the different types of vesicles in the metapopulation,
\begin{equation}\label{Phi}
\Phi_{t+1} \left ( \vec{m} \right ) = \sum_{\vec{k}} {}^\prime \Phi_{t} \left ( \vec{k} \right )
\left [ G_a \left ( \vec{m} \mid \vec{k} \right ) + G_b \left ( \vec{m} \mid \vec{k} \right )  \right ]
\end{equation}
where the primed sum is over viable vesicles, i.e., vesicles that contain at least one copy of each functional gene,
$k_l > 0$ for $l> 0$. This restriction is the expression of  the vesicle extinction process.  

In principle, the solution of the recursion equations (\ref{Phi}) yields detailed information on the
time evolution of the metapopulation. But the computational resources needed to generate
the entries of the matrices $G_a$ and $G_b$  seriously constrain the range of
$\Lambda$ and $d$ that can be studied in practice. In this contribution we show
how this difficulty can be circumvented by considering a finite population of vesicles 
with the same growth rate per generation
as the ideal, unrestricted metapopulation described above. Nevertheless, some interesting
information can be obtained from the analytical approach as described in Sec.\ \ref{sec:unrestricted}.

\subsection{Extinction probability} 

To point up the stochastic nature of the underlying vesicle dynamics -- the approach based on the 
recursion equations (\ref{Phi}) is deterministic as the focus is on the average number of a given vesicle type --
here we describe a general formulation to calculate the extinction probability $P_e \left ( \vec{k} \right )$
of the lineage sprouted by a vesicle of type $\vec{k}$. Generalizing 
the classic approach to evaluate the extinction probability in the Galton-Watson process \cite{Feller:68,Jagers:75}
we can write the following set of equations
\begin{equation}\label{ext_g}
P_e \left ( \vec{k} \right )= \sum_{\vec{m}_a,\vec{m}_b} P_{ab} \left ( \vec{m}_a, \vec{m}_b \mid \vec{k} \right ) 
P_e \left ( \vec{m}_a \right ) P_e \left ( \vec{m}_b \right )
\end{equation}
with the convention that $P_e \left ( \vec{m} \right ) = 1$ if the vesicle of type $\vec{m}$ is inviable. Note
that $P_e \left ( \vec{k} \right ) = 1, \forall \vec{k}$  is always a solution.
Surprisingly, this system of  $\mathcal{N}_V$  nonlinear coupled equations easily yields  to the simple iterative
solution method that begins with the guess  $P_e \left ( \vec{k} \right ) = 0$ for all viable vesicles.

%
\section{Unrestricted  growth}\label{sec:unrestricted}
%

It is clear from Eq.\ (\ref{Phi}) that the asymptotic regime of the  dynamics is
characterized either by the simultaneous  divergence or by the simultaneous
vanishing of $\Phi_{\infty} \left ( \vec{m} \right )$ for all viable vesicles. The main
goal here is to find the values of the control parameters $\Lambda$, $d$ and $u$ that
separates these two regimes. In general, this critical parameter setting can be found by direct numerical iteration
of the recursion equations.

Let us discuss first the simpler, extreme case $\Lambda=d$ for which there is only one type of  viable vesicle,
namely, $\vec{m}_* = \left ( 0,1,1, \ldots,1 \right)$.  In this case, Eq.\ (\ref{hyper}) simplifies considerably
and allows us to carry out analytically the summations in Eqs.\ (\ref{G_a}) and (\ref{G_b}). 
We find $G_a \left ( \vec{m}_* \mid \vec{m}_* \right ) = G_b \left ( \vec{m}_* \mid \vec{m}_* \right ) = \Omega_d (u)$
where
\begin{equation}\label{Omega}
\Omega_d \left ( u \right ) =  
\sum_{i=0}^d \left ( \begin{array}{c} d \\ i \end{array} \right )^2 \frac{i!}{d^i} \left ( 1 - u \right)^i 
 / \left ( \begin{array}{c} 2d \\ d \end{array} \right )
\end{equation}
and so  $\Phi_{t+1} \left ( \vec{m}_* \right ) = 2 \Omega_d \left ( u \right ) \Phi_{t} \left ( \vec{m}_* \right )$.
A straightforward numerical evaluation of Eq.\ (\ref{Omega}) for $u=0$  yields  $  \Omega_2 \left ( 0 \right ) = 7/12$
and  $  \Omega_{d>2} \left ( 0 \right ) < 1/2$. Since $\Omega_{d} $ is a monotonically decreasing function of $u$,
the latter inequality implies that $  \Omega_{d>2} \left ( u \right ) < 1/2$ for nonzero $u$ as well.  This 
indicates that template coexistence is unattainable for $\Lambda = d > 2$.   For   $\Lambda = d = 2$, 
however, the picture is different: 
the average size of the vesicle population will increase exponentially with increasing $t$ provided $u < u_c$
where $u_c = 3 - 2 \sqrt{2}$  is the solution of $\Omega_2 \left ( u \right ) = 1/2$. 
We find this simple analytical result reassuring because 
it proves that, even for finite vesicle capacities, functional templates can persist 
in the presence of a steady drain
towards the parasite class.

The evaluation of the extinction probability is also straightforward in the
case $\Lambda = d$.  Using Eq.\ (\ref{joint}) 
we write the probability that a viable vesicle produces two viable daughters (i.e., vesicles of
type $\vec{m}_*$) as
\begin{equation}
p_2 = d! \left [ \frac{1}{d} \left ( 1 - u \right ) \right ]^d ~ 
2^d /\left ( \begin{array}{c} 2d \\ d \end{array} \right ) .
\end{equation}
Now, Eq.\ (\ref{Omega}) yields the probability that  the first daughter of a viable vesicle is also viable, regardless
of the condition of the second daughter, so the probability that a viable vesicle produces a single viable offspring, no
matter whether it is the first or the second daughter,  is $p_1 = 2 \left ( \Omega_d  - p_2 \right )$. Hence 
the probability of producing two inviable daughters is $p_0 = 1 - p_1 - p_2$. In this case Eq.\ (\ref{ext_g}) reduces
to the simple quadratic equation $P_e = p_0 + p_1 P_e + p_2 P_e^2$ with $P_e = P_e \left ( \vec{m}_* \right )$.
The two solutions are $P_e = 1$ and $P_e = \left ( 2 \Omega_d - 1 \right )/p_2$.
The latter is physical provided $ \Omega_d > 1/2$ which, as pointed out before, holds only for $d=2$ and 
$u < 3 - 2 \sqrt{2}$.

For $\Lambda > d$, we have to resort to the numerical iteration of Eq.\ (\ref{Phi})
or to the numerical solution of Eq.\ (\ref{ext_g}) to obtain the critical parameter setting that determines the 
regime of viability of the metapopulation. In the former method, we begin the iteration ($t=0$) with a single 
parasite-free vesicle, whose composition of functional templates is as balanced as possible, 
e.g., $ \left( 0,\Lambda/d, \ldots, \Lambda/d \right )$ in the case of integer $\Lambda/d$. 

\begin{figure}
\centerline{\epsfig{width=0.52\textwidth,file=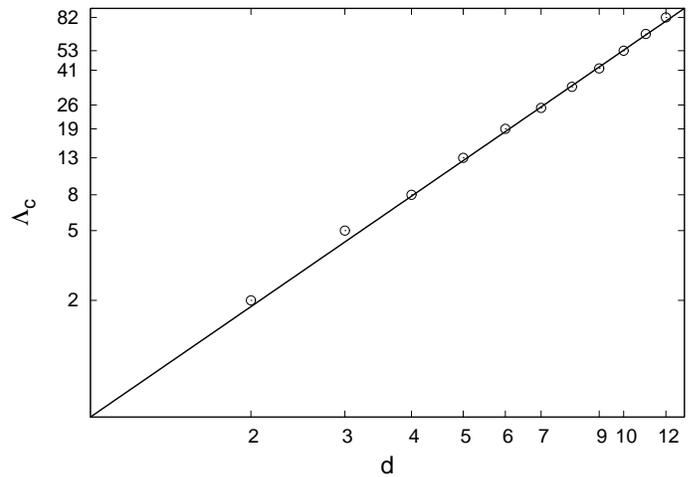}}
\par
\caption{Logarithmic plot of the value of the processivity $\Lambda_c$ above which the 
population size diverges against the number of functional templates $d$ in the case of
error-free replication $u=0$.
Below  $\Lambda_c$ the extinction of the lineage is certain. The solid  line is the fitting 
$\Lambda_c=d^2/2$. }
\label{fig:1}
\end{figure}

The results for the case of perfect replication accuracy $u=0$ are summarized in Fig.\ \ref{fig:1}.  
The critical processivity value $\Lambda_c$ above which the size of the metapopulation diverges
is very well described by the fitting  $\Lambda_c=d^2/2$  as shown in the figure. This
indicates that the assortment load can be compensated for if the redundancy $\Lambda/d$   
is larger than half the diversity value, i.e., provided that each vesicle contains at least $d/2$ copies of each 
functional template. This  simple result shows that there is no fundamental impediment to the coexistence of 
an arbitrary number of template types in the case of error-free replication if the cost of redundancy is
neglected. To understand the scaling $\Lambda_c \sim d^2$ at the critical boundary in the error-free replication
limit we must look at the ratio $r$ between the number of viable vesicles 
$\left ( \begin{array}{c} \Lambda - 1\\ d - 1 \end{array} \right )$
and the total number of vesicles $\left ( \begin{array}{c} \Lambda + d - 1\\ \Lambda \end{array} \right )$,
given by
\begin{equation}\label{r}
r =  \prod_{i=1}^{d-1}\frac{ 1 - i/\Lambda}{1 + i/\Lambda } .
\end{equation}
We note that since the parasite class is not taken into account in this error-free replication analysis, the number of
viable vesicles as well as the total number of vesicles differ from the quantities $\mathcal{N}_V$ and $\mathcal{N}_T$ introduced
before. The only way to obtain  nontrivial   values of  this ratio (i.e., $r \neq 0,1$) for large $\Lambda$ and $d$
is to suppose that  $d^2/\Lambda$ remains of order of 1. In this case we find $r \sim \exp \left ( - d^2/\Lambda \right )$ and so
$r_c = \mbox{e}^{-2} \approx 0.135$ at the critical boundary.

\begin{figure}
\centerline{\epsfig{width=0.52\textwidth,file=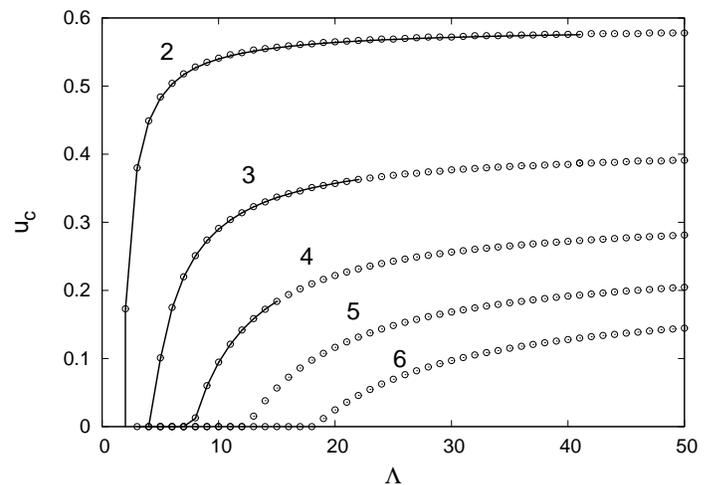}}
\par
\caption{Mutation probability $u_c$ above which the metapopulation is inviable  as a function of
the replicase processivity $\Lambda$ for  template diversity $d$ as
indicated in the figure. The lines are the analytical results for
unrestrained growth and the symbols are the results of the finite-size scaling analysis.}
\label{fig:2}
\end{figure}

We turn now to the analysis of the case where parasites are allowed, i.e., $u>0$. Fig.\ \ref{fig:2} summarizes the
main results, namely, the dependence on  $\Lambda$ and $d$ of the critical  mutation probability $u_c$ 
above which the lineage is inviable. The curves intersect the axis  $u_c=0$ at the values of $\Lambda$ exhibited in
Fig.\ \ref{fig:1}. The remarkable result revealed in Fig.\ \ref{fig:2} is that, for fixed $d$, there  exists a 
value of the mutation probability above which coexistence between the $d$ functional templates is impossible
regardless of the replicase processivity value  or, equivalently, the redundancy value. 
This result follows from the fact that $u_c$ tends to a well-defined
value less than 1 in the limit $\Lambda \to \infty$. This is reminiscent of the error threshold transition of the quasispecies
model for which the replication fidelity limits the length of the templates and hence the amount of information
that can be stored in the molecular population  \cite{Eigen:71}. Here the limitation is on the number of different
types of functional templates that can coexist within a vesicle and so on the total amount of information
that can be stored in the vesicle. 
Another result shown in Fig.\ \ref{fig:2} is the impracticability of the analytical approach for large $\mathcal{N}_T$:
the time required to evaluate the matrix entries defined in Eqs.\ (\ref{G_a}) and (\ref{G_b}) is simply prohibitive
so the curves are truncated at the values of $\Lambda$ that surpass  our computational resources. Fortunately, the
analysis of a finite population can greatly extend these limits, as we will show in Sec.\ \ref{sec:finite}.

\begin{figure}
\centerline{\epsfig{width=0.52\textwidth,file=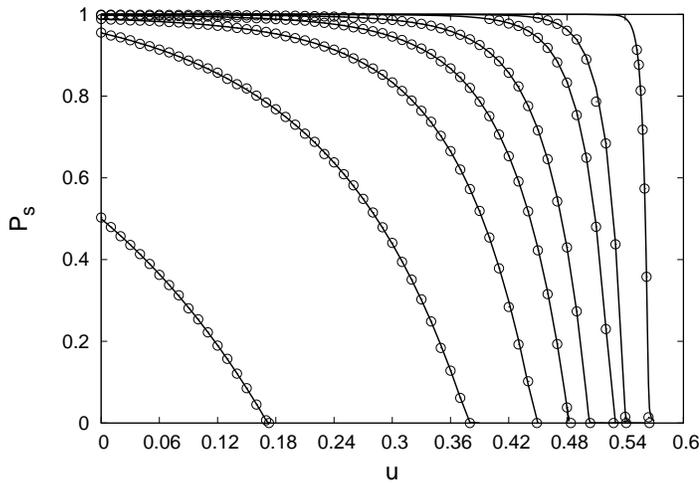}}
\par
\caption{Survival probability  of the lineage produced
by a single balanced, parasite-free vesicle
as function of the mutation probability $u$ 
for $d=2$ and (left to right) $\Lambda =2,3,4,5,6,8,10$, and $20$.
The lines are the analytical results for
unrestrained growth and the symbols are the results of the finite-population simulations with $N=10^3$
and $10^5$ independent samples.}
\label{fig:3}
\end{figure}

In addition to the threshold values exhibited in the previous figures, the analytical approach allows us
to obtain some detailed information  about the composition of the metapopulation and the nature of the stochastic process.
In particular, in Fig.\ \ref{fig:3} we show the survival probability $P_s = 1 - P_e$ of the lineage sprouted
by a  balanced, parasite-free vesicle obtained by solving numerically
Eq.\ (\ref{ext_g}) for  $d=2$. For $\Lambda = 3$ and $5$  
the ancestor vesicle is  $\left ( 0, \frac{\Lambda + 1}{d}, \frac{\Lambda - 1}{d} \right )$. Since the template 
dynamics becomes deterministic  in the limit 
$\Lambda \to \infty$, the survival probability must tend to a step function as indicated in the figure.

\begin{figure}
\centerline{\epsfig{width=0.52\textwidth,file=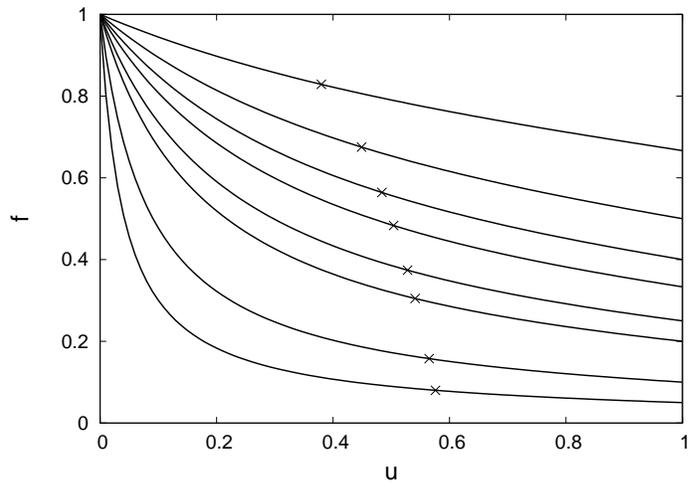}}
\par
\caption{Fraction of functional templates per viable vesicle in  the
steady-state regime for $d=2$ and (top to bottom) $\Lambda =3,4,5,6,8,10, 20$, and $40$. The symbol $\times$ indicates
the value of $u$ beyond which the population is inviable.}
\label{fig:4}
\end{figure}

Interestingly, we find that regardless of whether the process is subcritical ($u \ge u_c$) or supercritical 
($u < u_c$) the 
fraction $f$ of functional templates per viable vesicle, and consequently  the fraction of parasites, rapidly tends to a
steady-state  value. This fraction, defined by
\begin{equation}\label{f}
f = \lim_{t \to \infty} \frac{\sum_{\vec{k}} \left ( k_1 + \ldots + k_d \right ) \Phi_t  \left ( \vec{k} \right )}
{\Lambda \sum_{\vec{k}} \Phi_t  \left ( \vec{k} \right )}, 
\end{equation}
is shown in Fig.\ \ref{fig:4} as a function of the mutation probability.  For large $\Lambda$   we find that 
$f$ vanishes as $\Lambda^{-1}$ provided $u$ is nonzero and, in particular, $\Lambda f = d =2$ for $u=1$. So
in this limit 
there is only a finite number of functional templates in each vesicle, whereas the number of parasites grows
linearly with $\Lambda$.
The reason the population is viable (see Fig.\ \ref{fig:3}) even in these circumstances is that the
chances for choosing   functional templates for replication in some vesicle  is not 
negligible when the number of vesicles is greater than $\Lambda$,
which is always the case in the supercritical regime after a few generations.

%
\section{Finite population}\label{sec:finite}
%

As pointed out before, the impracticability of generating the entries of the matrices that govern the transitions between
viable vesicles for large values of $\Lambda$ and $d$  limits the applicability of the analytical solution summed up in
Eqs.\ (\ref{Phi}) and (\ref{ext_g}). To get around this obstacle we consider here an alternative
approach based on the Monte Carlo simulation of a finite population. In contrast to classical models
of populations genetics (e.g., the Wright-Fisher and Moran models \cite{Ewens:04})  in which the population size is kept fixed,
here we allow the number of vesicles to vary from $0$ to a fixed maximum value $N$. The idea is to implement the dynamics
exactly as done in the case of unrestrained growth, except that whenever the number of vesicles becomes greater than $N$, the surplus vesicles
are discarded randomly. The discard takes place before the check of the viability of the vesicles (extinction
process). This scheme is reminiscent of the so-called Russian Roulette used in the Monte Carlo simulation of
neutron production in nuclear reactors \cite{Hammersley:64}. 

In the subcritical  regime, the  introduction of
the  upper bound $N$ is innocuous since the population size is likely to remain small   before the 
extinction outcome anyway. In the supercritical regime, however, a too small bound may
prevent the lineage to produce and retain a minimum number of viable vesicles that would avoid extinction and so
one expects  $u_c (N) < u_c$. (An operational definition of $u_c(N)$ will be given later.) The finite
population scheme  is effective in the practical situation  $N \ll \mathcal{N}_V$ 
provided that only a small fraction of the $\mathcal{N}_V$ viable vesicles would actually be present in 
the metapopulation if it were allowed to grow  unrestrained.

\begin{figure}
\centerline{\epsfig{width=0.52\textwidth,file=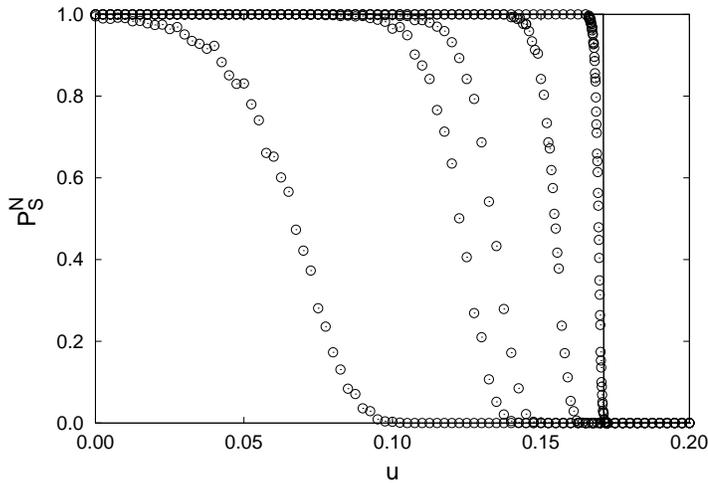}}
\par
\caption{The survival probability of a population composed initially by  $m=N$ 
balanced, parasite-free  vesicles for $\Lambda = d=2$ and  (left to right) $N= 40, 80, 100, 200 $ and $1000$.
The solid line is the analytical result for unrestrained growth. Each symbol represents an average
over $10^5$ samples.}
\label{fig:5}
\end{figure}

\begin{figure}
\centerline{\epsfig{width=0.52\textwidth,file=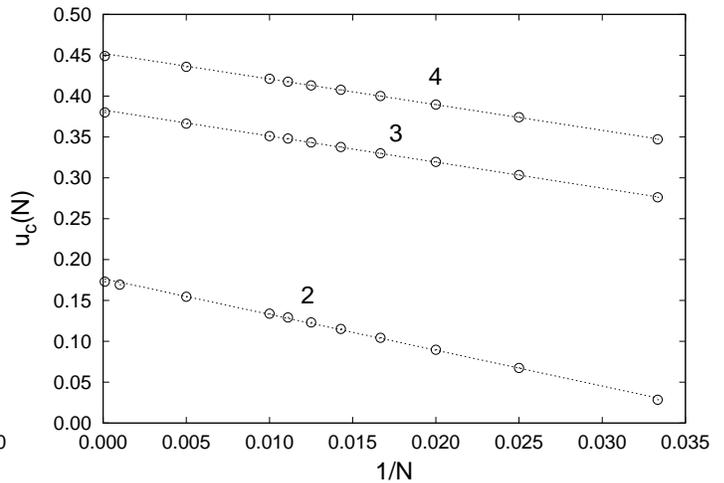}}
\par
\caption{Mutation probability $u_c(N)$ at which $50\%$ of  samples of a population of $N$ balanced, parasite-free 
vesicles survive for $d=2$  and $\Lambda = 2,3$, and $4$. The linear fittings (dashed lines) 
$u_c(N) = u_c - a_\Lambda/N$ allow us to estimate $u_c$. }
\label{fig:6}
\end{figure}

The previous setup for the initial structure of the population -- a single balanced, parasite-free vesicle -- is not
suited to study  finite size effects on the estimate of the critical mutation probability, because there is
no operational way to define $u_c (N)$. For that end an effective strategy is
to begin with a population of $m$ identical such  vesicles. Since the vesicles evolve independently
there is a simple relationship between the probability that a population with initial size $m$ thrives, denoted by $P_s^m$,
and the survival probability of a single vesicle $P_s$ exhibited in Fig.\ \ref{fig:3}, namely,
\begin{equation}\label{Ps_N}
P_s^m = 1 - \left ( 1 - P_s \right )^m .
\end{equation}
For $m \to \infty$ this quantity tends to a step function that takes on the values $1$ if $u < u_c$ and
$0$ otherwise. In the finite population simulations we set $m=N$, since our focus in on the behavior of
$P_s^m$ when  both quantities - the
initial size $m$ and the size upper limit $N$ -  become arbitrarily large.
The results for $P_s^N$ are shown in Fig.\ \ref{fig:5}
for different population sizes $N$. As $N$ increases, the finite-population results approximate those for 
the unrestrained growth represented by the step function. To quantify this approach, we arbitrarily  define 
$u_c \left ( N \right )$ as the value
of the  mutation probability at which $P_s^N = 1/2$ so that  the critical
value $u_c$ for $N \to \infty$ can be inferred as illustrated in Fig.\ \ref{fig:6}. 

Use of $P_s^N$ instead of $P_s$ is crucial for this analysis, since regardless of the definition of
$u_c(N)$ (e.g., we could define it as the mutation probability at which $P_s^N = x$ for any $0 < x < 1$) 
this quantity tends to $u_c$ in the limit of $N$ large.
The extrapolated values to $1/N \to 0$ are presented in Fig.\ \ref{fig:2} and agree perfectly
with the available analytical predictions. This gives us confidence to use the finite population
estimates in the cases where the analytical approach is not practical.

\begin{figure}
\centerline{\epsfig{width=0.52\textwidth,file=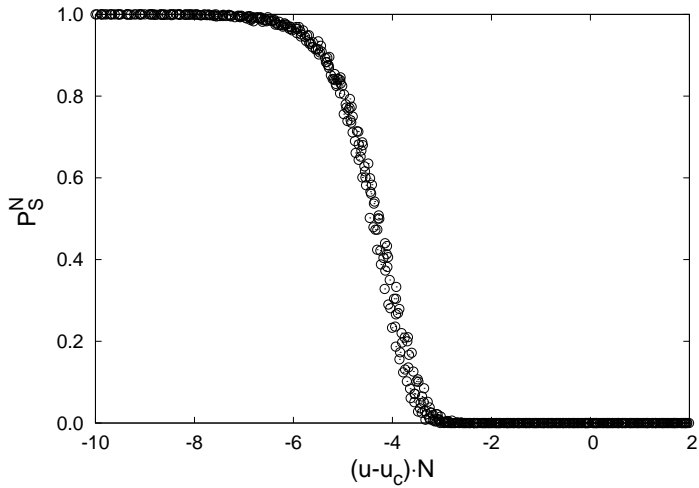}}
\par
\caption{The data  of Fig.\ \ref{fig:5} ($d=\Lambda = 2$) plotted against the scaled  mutation probability
$\left ( u - u_c \right ) N$ with $u_c = 3 - 2 \sqrt{2}$.
The collapse of the data into a single $N$-independent function signals the occurrence of
a threshold phenomenon   at $u_c$. }
\label{fig:7}
\end{figure}

Fig.\ \ref{fig:7} summarizes the results of the data collapsing method (see, e.g., \cite{Barber:83,Binder:85})
applied to the data of Fig.\ \ref{fig:5}. The survival probabilities $P_s^N$ collapse into
a single universal form (scaling function) if plotted against the scaled mutation probability
$\left (u-u_c \right) N$ where $u_c$ is the critical mutation probability of the infinite population.
This scaling function shows that, for $u-u_c$ fixed, $\lim_{N \to \infty} P_s^N$ tends to 1 if $u < u_c$ and
to $0$ otherwise, and that the characteristics of the threshold transition persist across a range of $u$ of order
$1/N$ about $u_c$. A similar finite-size scaling analysis was employed to fully characterize the error threshold transition of
the quasispecies model in the cases where the population size as well as the molecules lengths are fixed and finite 
\cite{Campos:98}.

Of course, the finite population scheme can be used to study the original setup in which the initial
population comprises a single vesicle, $m=1$, as well. In this case the size limit $N=1000$
suffices to obtain perfect agreement with the analytical results, as shown in Fig.\ \ref{fig:3}.

%
\section{Discussion}\label{sec:Disc}
%

\begin{figure}
\centerline{\epsfig{width=0.52\textwidth,file=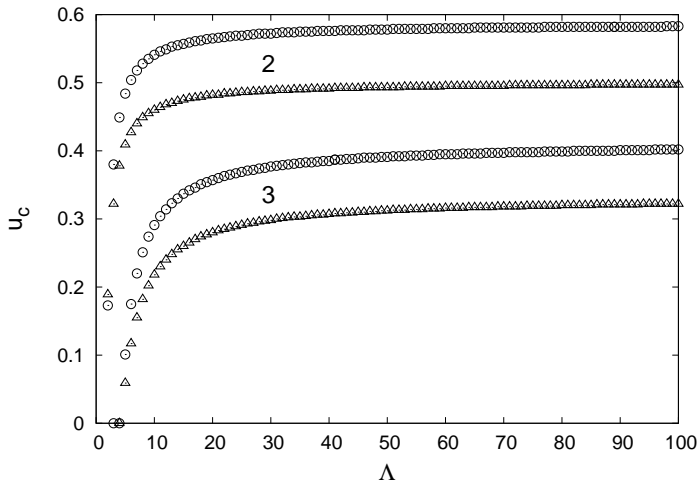}}
\par
\caption{Critical mutation probability obtained via finite-size scaling for the
package model proposed by Niesert et al.\  ($\triangle$) and the variant proposed in this paper
($\bigcirc$) for $d=2$ and $3$ as indicated. }
\label{fig:8}
\end{figure}

In Sec.\ \ref{sec:model} we have outlined the differences between the original package model
proposed  by Niesert et al.\ and our more tractable variant. Here we present a brief comparison
between the main predictions of these models. The first important observation is that in the
absence of parasites ($u=0$) both models yield identical results for the viability boundary
in the plane $(d,\Lambda)$  (see Fig. \ref{fig:1}). When parasites are present, however, there is
a substantial quantitative difference between the critical mutation probability of the two models,
as shown in Fig.\ \ref{fig:8}. The   scheme based on synchronous  template replication
and symmetric fission seems to be considerably more robust to the action of parasites than
the less structured procedures of the original  proposal. This result supports the view that
the mechanisms of segregation of modern cells originated in response to mutation pressure
\cite{Santos:98}.

It is hard to see why parasites are more harmful in the asynchronous replication and
asymmetric fission setting. Considering  that parasites are rare at the beginning, their
spread should be hampered by the asynchronous replication scheme in the initial generations and then
speeded up when the parasites become  numerous. By simulating the two template replication schemes 
with the same fission mechanism,  we have verified
that the choice of the form of update -- parallel or sequential -- practically
does not affect the critical mutation probability $u_c$. Thus the key element to explain
the quantitative differences between the models illustrated in Fig.\ \ref{fig:8} must
be the fission mechanism. To get some insight on that, let us consider the situation where a mother
vesicle of size $2 \Lambda$ contains  two  functional templates of a certain type. 
Clearly, from the mother vesicle's perspective the optimal strategy is to send  one template to each 
of her daughters. The probability this happens for the asymmetric fission strategy is $1/2$ independently
of the vesicle size. The symmetric fission scheme in turn yields a slightly larger  probability for
this event, namely,  $1/2 \times  \left ( 1 - 1/2 \Lambda \right)^{-1}$. This tendency of the symmetric fission strategy 
to a more balanced  distribution of templates of the same type to the daughters is probably the reason of its enhanced
robustness against parasites.

Our finding that $u_c$ is a nondecreasing function of $\Lambda$ (see Fig.\ \ref{fig:8}) is at variance with
the results of Niesert et al.\ which predict that $u_c$ would reach a maximum 
and then decrease towards zero as  $\Lambda$  increases further \cite{Niesert:81}. The reason may be  the
criterion for discard of supernumerary vesicles used in that work, which was based on  three properties:  the degree of equipartition 
of the copies among the 
different functional templates, the number of parasites and the overall redundancy of the functional templates. 
In fact, we have verified (see Fig.\ \ref{fig:4}) that for large $\Lambda$ and not too low $u$, 
the surviving vesicles in the supercritical regime are heavily loaded with parasites and so use of such 
selection criterion would purge them from the population resulting  in a premature extinction.

Although the finite population simulations were used here as a tool to validate and complement
the analytical results, they are of interest on their own. In particular, the
Muller ratchet \cite{Muller:64,Felsenstein:74} and the mutational meltdown \cite{Lynch:93} 
are important stochastic phenomena that result in the accumulation of mutations  in finite populations (see \cite{Fontanari:03}
for the study of both phenomena in growing lineages). In our framework, the counterpart of accumulation of mutations  is
the accumulation of inviable vesicles, which is explicitly ruled out by the assumption that those vesicles
are unable to divide into daughter vesicles. This peculiar aspect of the model was severely  criticized by Eigen et al.
\cite{Eigen:80} who pointed out that the vesicles in the model Niesert et al.  \cite{Niesert:81} cannot evolve  
because of that assumption. Since there is no competition among the vesicles in the case of unrestricted growth, allowing
the inviable vesicles to divide as well would have no effect at all on the dynamics of the viable vesicles because it is
not possible to produce a viable vesicle by fissioning an inviable one. 
In the finite population case, on the other hand, the inviable vesicles would accumulate steadily and ultimately
would reach fixation in the metapopulation.

\section{Conclusion}

The goal of the research on prebiotic evolution is to put forth a coherent scenario for
the origin and early development of life. So at this stage it is appropriate to appraise
the main results of our analysis of this classic package model, summarized in Fig.\ \ref{fig:2}.
Given the spontaneous error rate 
per nucleotide $\epsilon$ and the molecule length $L \gg 1$ we can readily obtain the value of the 
probability  of mutation from functional templates to parasites,
 $u = 1 - \exp \left ( - \epsilon L \right )$.
A plausible estimate for these primary parameters is $\epsilon \sim 10^{-2}$ and
$L \sim 100$ \cite{Eigen:71} which yields $u \sim 0.6$. A glance at Fig.\ \ref{fig:2} leads
to the disastrous conclusion that even the coexistence between two templates is prohibited in
these circumstances. It is instructive also  to compare our results with those of the hypercycle 
 which guarantees the stable coexistence of at most  
$d=4$ templates \cite{Eigen:78} (see \cite{Campos:00} for the analysis of the hypercycle in the presence of an error tail class
similar to the parasite class considered here). According to Fig. \ref{fig:2}, $d=4$ functional templates 
can coexist provided that $u < 0.3$ which implies that $L < 35$, resulting in  a total of $140$ nucleotides,
a meager improvement over the $100$ nucleotides prediction for a free replicator. 

It should be observed, in addition, that
the critical mutation probabilities $u_c$ exhibited in Figs. \ref{fig:2} and \ref{fig:8}
are  best case results since neither lethal mutations nor accidents were considered in our calculations. Hence,
contrary to the claims of Niesert et al.\ \cite{Niesert:81}, the kind of template coexistence achieved in 
a simple package model does not resolve the prebiotic information crisis.  Special
mechanisms to prevent independent information carriers from competing with one another within the compartment
must be posited. Although unwarranted assumptions like  the hypercyclic organization \cite{Eigen:78} or the coupling between the
templates and the package metabolism \cite{Czaran:00,Zintzaras:02,Silvestre:05,Fontanari:06}
may weaken the credibility of the models, so far they seem to stand as the only options to tackle
the  information crisis of prebiotic evolution.

\begin{acknowledgments}
This research  was supported by CAPES, CNPq and FAPESP, Project No. 04/06156-3.
\end{acknowledgments}


\end{document}